\documentstyle[aps,prd,tighten,epsf]{revtex}
\begin{document}
\draft
\newcommand{\be}{\begin{equation}}
\newcommand{\ee}{\end{equation}}
\newcommand{\ben}{\begin{eqnarray}}
\newcommand{\een}{\end{eqnarray}}

\newcommand{\la}{{\lambda}}
\newcommand{\Om}{{\Omega}}
\newcommand{\ta}{{\tilde a}}
\newcommand{\bg}{{\bar g}}
\newcommand{\bh}{{\bar h}}
\newcommand{\si}{{\sigma}}
\newcommand{\th}{{\theta}}
\newcommand{\C}{{\cal C}}
\newcommand{\D}{{\cal D}}
\newcommand{\cA}{{\cal A}}
\newcommand{\cT}{{\cal T}}
\newcommand{\cO}{{\cal O}}
\newcommand{\eeo}{\cO ({1 \over E})}
\newcommand{\G}{{\cal G}}
\newcommand{\cL}{{\cal L}}
\newcommand{\T}{{\cal T}}
\newcommand{\M}{{\cal M}}

\newcommand{\p}{\partial}
\newcommand{\na}{\nabla}
\newcommand{\ssum}{\sum\limits_{i = 1}^3}
\newcommand{\dssum}{\sum\limits_{i = 1}^2}
\newcommand{\tal}{{\tilde \alpha}}

\newcommand{\tp}{{\tilde \phi}}
\newcommand{\tPhi}{\tilde \Phi}
\newcommand{\tpsi}{\tilde \psi}
\newcommand{\tim}{{\tilde \mu}}
\newcommand{\tr}{{\tilde \rho}}
\newcommand{\tir}{{\tilde r}}
\newcommand{\rp}{r_{+}}
\newcommand{\hr}{{\hat r}}
\newcommand{\rv}{{r_{v}}}
\newcommand{\dr}{{d \over d \hr}}
\newcommand{\dR}{{d \over d R}}

\newcommand{\hhf}{{\hat \phi}}
\newcommand{\hhM}{{\hat M}}
\newcommand{\hhQ}{{\hat Q}}
\newcommand{\hht}{{\hat t}}
\newcommand{\hhr}{{\hat r}}
\newcommand{\hhS}{{\hat \Sigma}}
\newcommand{\hhD}{{\hat \Delta}}
\newcommand{\hhm}{{\hat \mu}}
\newcommand{\hro}{{\hat \rho}}
\newcommand{\hhz}{{\hat z}}

\newcommand{\tD}{{\tilde D}}
\newcommand{\tB}{{\tilde B}}
\newcommand{\tV}{{\tilde V}}
\newcommand{\hT}{\hat T}
\newcommand{\tF}{\tilde F}
\newcommand{\tT}{\tilde T}
\newcommand{\hC}{\hat C}
\newcommand{\ep}{\epsilon}
\newcommand{\bep}{\bar \epsilon}
\newcommand{\ppp}{\varphi}
\newcommand{\Ga}{\Gamma}
\newcommand{\ga}{\gamma}
\newcommand{\hth}{\hat \theta}
\title{Late-Time Evolution of Charged 
Massless Scalar Field in
the Spacetime of Dilaton Black Hole}

\author{ Rafa{\l} Moderski}
\address{JILA, University of Colorado \protect \\
CB 440, Boulder, CO 80309-0440 \protect \\
and \protect \\
Nicolaus Copernicus Astronomical Center \protect \\
Polish Academy of Sciences \protect \\
00-716 Warsaw, Bartycka 18, Poland \protect \\
moderski@camk.edu.pl }

\author{Marek Rogatko}
\address{Technical University of Lublin \protect \\
20-618 Lublin, Nadbystrzycka 40, Poland \protect \\
rogat@tytan.umcs.lublin.pl \protect \\
rogat@akropolis.pol.lublin.pl}
\date{\today}
\maketitle
\smallskip
\bigskip
\begin{abstract}
We investigate the power-law tails in the evolution of a charged
massless scalar field around a fixed background of dilaton black
hole. Using both analytical and numerical methods we find the inverse 
power-law relaxation of charged fields at future timelike infinity,
future null infinity and along the outer horizon of the
considered black hole. We invisaged that a charged hair decayed slower
than neutral ones. The oscillatory inverse power-law along the outer
horizon of dilaton black hole is of a great importance for a mass
inflation scenario along the Cauchy horizon of a dynamically formed
dilaton black hole.
\end{abstract}
\baselineskip=18pt
\par
\section{Introduction}

The late-time evolution of various fields outside a collapsing
star plays an important role in major aspects of black hole
physics, i.e., in the Wheeler's {\it no-hair theorem} \cite{bh}
and in {\it mass inflation} scenario \cite{mi}.
In Ref.\cite{pri}
the neutral external perturbations were studied and it
was found that the late-time behaviour for a fixed $r$ is dominated by
the factor $t^{-(2l + 3)}$, for each multipole moment $l$. Next,
Gundlach {\it et al} \cite{gun} studied the behaviour of neutral 
perturbations along null infinity and along future event horizon. Their
conclusion was that along null infinity the decay is according the power 
law of the form 
$u^{-(l+2)}$, where $u$ is the outgoing Eddington-Filkenstein (ED)
null coordinate. On the other hand, the neutral perturbations along the 
event horizon behave like $v^{-(l+3)}$, where $v$ is the ingoing ED
coordinate. Bicak \cite{bic} analysed the scalar field perturbations on
Reissner-Nordstr\"om (RN) background and found for $\mid Q \mid < M$
the relation $t^{-(2l +2)}$, while for $\mid Q \mid = M$ the late-time
behaviour for a fixed $r$ was governed by the rule $t^{-(l +2)}$.
\par
The late-time behaviours of a charged massless scalar field in
RN spacetime were studied analytically and confirmed by the direct 
numerical studies of the fully nonlinear gravitational collapse
in \cite{pir1,pir2,pir3}.
Among all, the conclusion was that a charged hair decayed slower than a
neutral one, i.e., the charged scalar hair outside a charged black hole
was dominated by a $t^{-(2l +2)}$ behaviour.
\par
The problem of the late-time tails in gravitational collapse of a self
interacting massive field were studied in \cite{mm} (see also
\cite{bur,ja,ja1}).
At late-times the decays of a self-interacting hair is 
slower than any power law.
\par
In our work we shall discuss the asymptotic evolution of a massless
charged scalar field in the background of dilaton black hole.
Focusing on {\it no-hair theorem} we shall be interested in the
dynamical mechanism by which the charged hair is radiated away.
In Sec.II we start with the outline of the system under consideration
and we analitycally study the late-time evolution of charged scalar
perturbations along timelike infinity, future null infinity
and the black hole outer horizon. Then, in Sec.III we treated  the
problem numerically. Sec.IV outlines our conclusions and remarks.

\section{The Einstein-Maxwell-dilaton Equations}
\label{sec1}
In this section 
we shall analytically study the evolution of a massless charged scalar 
field $\psi$ around a fixed
background of electrically charged
dilaton black hole. The wave Eq. for a massless
charged scalar field is given by \cite{haw}
\be
\na_{\mu} \na^{\mu} \psi - i e A_{\alpha} g^{\alpha \beta}
(2 \na_{\beta} \psi - i e A_{\beta} \psi) - ie \na_{\mu} A^{\mu} \psi = 0,
\ee
where $e$ is a constant value and $A_{\beta}$ is the gauge potential.
\par
The metric of the external gravitational field will be given by
the static, spherically symmetric solution of Eqs. of motion derived from
the low-energy string action (see e.g.\cite{dbh}). The action has the form
as follows:
\be
S = \int d^4 x \left [
R - 2 (\na \phi)^2 - e^{-2 \phi} F^2 \right ],
\ee
where $\phi$ is the dilatonic field and $F_{\alpha \beta} =  2\na_{[
\alpha} A_{\beta ]}$.
The metric of the electrically charged dilaton black hole is as follows:
\be
ds^2 = - \left ( 1 - {2 M \over r} \right )dt^2 +
{dr^2 \over \left ( 1 - {2 M \over r} \right )} +
r \left ( r - {Q^2 \over M} \right ) (d\theta^2 + \sin \theta d \phi^2 ).
\label{met}
\ee
The event horizon is located at $r_{+} = 2 M$. For the case of 
$r_{-} = {Q^2 \over M}$ we have another singularity but it can be ignored 
because of the fact that $r_{-} < r_{+}$. The dilaton field is given by 
$e^{2 \phi} = e^{- 2 \phi_{0}} \left ( 1 - {r_{-} \over r} \right )$, 
where
$\phi_{0}$ is the dilaton's value at $r \rightarrow \infty$. The mass
$M$ and the charge $Q$ are related by the relation $Q^2 = {r_{+}r_{-}
\over 2} e^{2 \phi_{0}}$.\\
Defining the tortoise coordinate $y$, as
$dy = {dr \over \left ( 1 - {2 M \over r} \right ) }$ one can
rewrite the metric (\ref{met}) in the form
\be
ds^2 = \left ( 1 - {2 M \over r} \right ) \left [
- dt^2 + dy^2 \right ] + r \left ( r - {Q^2 \over M} \right ) 
(d\theta^2 + \sin \theta d \phi^2 ).
\ee
For the spherical background each of the multipole of perturbation
field evolves separetly so for the scalar field in the form
$\psi = \sum_{l,m} \eta_{m}^{l}(t, r) Y_{l}^{m}(\theta, \phi)/R(r)$, 
one has the 
following Eqs. of motion for each multipole moment
\be
\eta_{,tt} - 2 ie A_{t} \eta_{,t} - \eta_{,yy} + V \eta = 0,
\label{mo}
\ee
where
\be
V = \left [
{l(l + 1) \over R^2} + {R'' \over R} \left ( 1 - {2 M \over r} \right )
+ {2 M R' \over r^2 R} \right ] \left ( 1 - {2 M \over r} \right )
-e^2 A_{t}^{2}.
\ee
By $R$ we denoted $R = \sqrt{r \left ( r - {Q^2 \over M} \right )}$
and $'$ is the derivative with respect to the $r$-coordinate.
As in Ref.\cite{pir1} in order 
to get rid of the physically unimportant quantity $\tPhi$,
which appear in $A_{t} = \tPhi - {Q \over r}$,
one can define the auxilary field $\tpsi = e^{- ie \tPhi t} \eta$.
Then Eq.(\ref{mo}) may be written as
\be
\tpsi_{,tt} + 2ie{Q \over r} \tpsi_{,t} - \tpsi_{,yy} +
\tV \tpsi = 0,
\label{mmo}
\ee
where
\be
\tV = \left [
{l(l + 1) \over R^2} + {R'' \over R} \left ( 1 - {2 M \over r} \right )
+ {2 M R' \over r^2 R} \right ] \left ( 1 - {2 M \over r} \right )
- {e^2 Q^2 \over r^2}.
\label{mm}
\ee
One can assume that the general solution of Eq.(\ref{mmo}) is of the form
\ben
\tpsi &=& \sum_{k = 0}^{l} A_{k} R^{- k}
\left [
e^{- ie Q ln r} G(u)^{(l - k)} + (-1)^{k} e^{ie Q ln r} F(v)^{(l - k)}
\right ] \\ \nonumber
&+& \sum_{k = 0}^{\infty} \left [
B_{k}(R) G(u)^{(l - k - 1)} + C_{k} F(u)^{(l - k -1)} \right ],
\een
where $F$ and $G$ are arbitrary functions
of an retarded time coordinate $u = t - y$ and an advanced time
coordinate $v = t + y$. The superindices on $F$ and
$G$ 
have different meaning depending on negativity or positivity of them.
The positive superindices indicate the number of times the function is
differentiated, while the negative ones are to be interpreted as
integrals \cite{pri}. 
The first sum in $\tpsi$ represents the primary waves in the wavefront 
while the second sum depicts the backscattered waves.
The first sum in the above expression is the zeroth order
solution.\\
The recursion relation for $B_{k}(R)$ is as follows:
\ben \label{ba}
2 \la^2 B_{k}(R)' &+& {2 ieQ \over r} B_{k}(R) - \la^2 \left [
\la^2 B_{k - 1}(R)' \right ]' +
\la^2 \left [ A_{k}(R) (k R'+ ie Q) {1 \over R^{k + 1} r^{i e Q}} 
\right ]'  \\ \nonumber
&+& \tV \left [ A_{k}(R) R^{- k}r^{-ieQ} + B_{k - 1} (R)
\right ] + 2 A_{k + 1}(R) {\left [
ieQ (R - r \la^2) - \la^2 r R' (k + 1) \right ] \over
R^{k + 1} r^{ieQ + 1}} = 0,
\een
where $\la = \left ( 1 - {2 M \over r} \right )$.
We expand the $B_{k}(R)$ coefficient in the form
\be
B_{k}(R) = {a_{k} \over R^{k + 1} r^{ieQ}} +
{b_{k} \over R^{k + 1} r^{ieQ + 1}} + ...,
\label{bb}
\ee
where as in \cite{pri} $a_{k} = a_{k}(l, eQ)$, $b_{k} = b_{k}(l, eQ)$.\\
Using Eqs.(\ref{ba}) and (\ref{bb}) the lowest order coefficients yield
\be
a_{k} = - ieQ A_{k}(R) {2k +1 \over 2(k + 1)} \left [
1 + \cO(e Q)\right ].
\ee
As was pointed out after starting of collapsing of the star surface it 
approached very fast the speed of light. Thus its word line is asymptotic
to an ingoing null geodesics $v = v_{0}$. The time dilatation between
static frames and infalling frames caused that the variation of the field 
$\tpsi$
on the star's surface is asymptotically redshifted. This means that
$\tpsi_{,u}$ will be small at late times. Then it is well justified 
to make an assumption that after some retarded time $u = u_{1}$ variations
of the charged field $\tpsi$ on $v = v_{0}$ can be neglected.\\
In the first stage of evolution we shall consider the scattering 
in the region $u_{0} < u < u_{1}$. One has taken the variation of the field
on $v = v_{0}$ to be neglected after $u_{1}$. Thus, we have no outgoing 
radiation for $u > u_{1}$, except from backscattering. This
assumption about no late outgoing waves is equivalent to taking
$G(u_{1}) = 0$. Summing it all up, for large $r$ at $u = u_{1}$
the dominant term in $\tpsi$ (the dominant backscatter of the primary
waves) is as follows:
\be
\tpsi (u = u_{1}, r) = a_{l}{1 \over r^{l+1}} G^{(-1)}(u_{1}).
\label{t0}
\ee
In the next stage of the consideration we take a closer
look at the asymptotic evolution of the charged scalar field. We shall
take into consideration the region where $y \gg M, Q$ and 
for which $u > u_{1}$.
In this region the leading order effect on the propagation of $\tpsi$, 
has the {\it centrifugal term} in the potential $\tV$, namely the
evolution is dominated by a neutral flat spacetime terms. Thus
one has
\be
\tpsi_{,tt} - \tpsi_{,rr} + {l(l + 1) \over r^2} \tpsi = 0.
\label{t1}
\ee
As in RN case in this region one can replace $r$ by $y$, because
of the fact that $\tpsi_{,yy} \simeq \tpsi_{,rr} + \cO(M)$.
The solution to Eq.(\ref{t1}) can be written in the form 
\be
\tpsi = \sum_{l=0}^{k} A_{k}y^{-k} \left [
g^{(l-k)}(u) + (-1)^{k} f^{(l - k)}(v) \right ].
\label{t2}
\ee
Matching (\ref{t2}) with initial data on $u = u_{1}$ Eq.(\ref{t0}), one has
$f(v) = F_{0}v^{-1}$, where $F_{0}$ is of the same form as
in RN case \cite{pir1}. The standard procedure of
expansion \cite{pri}, for 
late times $t \gg y$ for $g(u)= \sum_{n=0}^{\infty}
{(-1)^{n} \over n! }g^{(n)}(t) y^n$ and similary $f(v) = 
\sum_{n = 0}^{\infty} {1 \over n!} f^{(n)}(t) y^n$, 
enables one to rewrite (\ref{t2}) as
follows:
\be
\tpsi = \sum_{n=-1}^{\infty} K_{n} y^{n} \left [
f^{(l + n)}(t) + (-1)^{n}g^{(l+n)}(t) \right ],
\label{t3}
\ee
where $K_{n}$ are given in Ref.\cite{pri} as in the neutral case.\\
Using the boundary conditions for small $r$ (see \cite{pri} 
and \cite{gun}
for the details), one finds that the 
late time behaviour of the scalar field at timelike infinity $i_{+}$ is
\be
\tpsi \simeq 2 K_{l+1} y^{l + 1} f^{(2l + 1)}(t) =
- 2 K_{l+1} F_{0}(2l + 1)!t^{-2(l + 1)}y^{l + 1} + \cO(e Q) .
\label{ip}
\ee
On a future null infinity $scri_{+}$ one has
\be
\tpsi(v \gg u, u) \simeq A_{0} g^{(l)}(u) \simeq - F_{0} l! u^{-(l+1)}.
\ee
At the black hole outer horizon $H_{+}$
(as $y \rightarrow - \infty$) the
Eq. of motion can be approximated by
\be
\tpsi_{,tt} + {ieQ \over M} \tpsi_{,t} - \tpsi_{,yy} - 
{e^2 Q^2 \over 4 M^2} \tpsi = 0,
\ee
with the general solution of the form
\be
\tpsi = e^{-{ieQ \over 2M}t} \left [ \alpha(u) + \ga(v) \right ]
= e^{-{ieQ \over 2M}t} \ga(v).
\label{ho}
\ee
We put $\alpha(u) = 0$ because of the condition of
neglecting the variation of the scalar field on $v = v_{0}$.
To join the solution for $y \ll - M$ with $y \gg M$ for $i_{+}$, 
one uses the ansatz \cite{gun,pir1}
$\tpsi \simeq \tpsi_{stat} t^{-2(l + 1)}$ for the solution in region 
$y \ll - M$ and $t \gg \mid y \mid$, i.e., we assume that for 
the whole range of $y$ the solution has the same late-time dependence
as for $y \gg M$. This enables us to match the solution (\ref{ho})
with that for $i_{+}$. One concludes that the late-time behaviour
on the horizon $H_{+}$ is
\be
\tpsi (u \rightarrow \infty, v) = \Ga_{0} e^{ ieQ y \over 2 M} v^{
-2(l + 1)},
\ee
where the constant $\Ga_{0}$ is detrmined by the condition 
that there is a static solution fulfilling $lim_{y \rightarrow \infty}
\tpsi_{finstat} = \Ga_{0}$ and $lim_{y \rightarrow \infty}
\tpsi_{finstat} = - 2 K_{l}^{2l + 1}F_{0}(1 + 2l)!  r^{l + 1 }$.
As was remarked in \cite{pri,gun}, the coefficient $\Ga_{0}$  and all 
other aspects of the late-time behaviour are governed by the 
initial backscattering.\\
As one can see the behaviour of the charged scalar field on 
a fixed dilatonic background at timelike null infinity $i_{+}$
and the future null infinity $scri_{+}$ is the same as in the
RN black hole case. However, on the external horizon $H_{+}$
we have some changes, but the general form of the time dependence
is maintained.

\section{Numerical method and results}
\begin{figure}[t]
\begin{center}
\leavevmode
\epsfxsize=220pt
\epsfbox{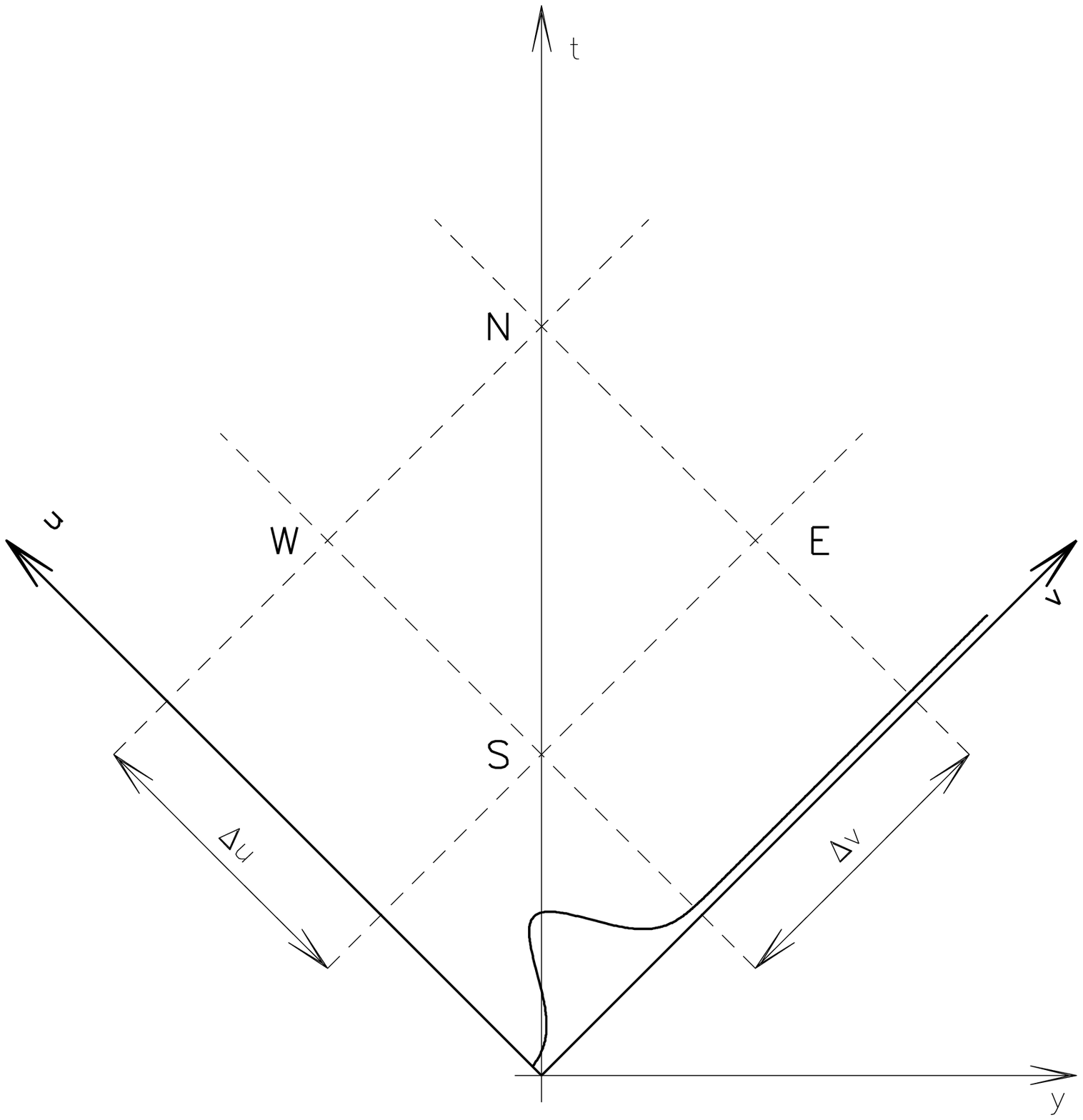}
\end{center}
\caption{Schemat of computational grid. Initial conditions are
specified on $u=u_0$ axis, and boundary conditions on $v=v_0$
axis. Given the values of the field at W,S and E points we can
calculate the field at N. Starting from lower corner of the grid
calculation proceeds to upper corner row by row.}
\label{fig1}
\end{figure}

In this section we shall numerically study Eqs.(\ref{mmo}) and
(\ref{mm}). First we transform Eq.(\ref{mmo}) into $(u,v)$
coordinates. This yields:
\be 
4 \tpsi_{,uv} + 2ie{Q \over r} (\tpsi_{,u} + \tpsi_{,v}) +
\tV \tpsi = 0. 
\ee
Then we divide this Eq. into a set of coupled equations for real,
$\tpsi_{\rm R}$, and imaginary, $\tpsi_{\rm I}$, parts of the field.
\ben
4 \tpsi_{{\rm R},uv} - 2e{Q \over r} (\tpsi_{{\rm I},u} + \tpsi_{{\rm I},v}) +
\tV \tpsi_{\rm R} = 0. \\
4 \tpsi_{{\rm I},uv} + 2e{Q \over r} (\tpsi_{{\rm R},u} + \tpsi_{{\rm R},v}) +
\tV \tpsi_{\rm I} = 0.
\een
To solve these equations numerically we establish uniformally spaced
grid $(u_i,v_i)$, $i=0,...,n$, with $u_i=v_i=i h$ and $\Delta u =
\Delta v = h$. Next, we construct implicit difference scheme:
\ben
\tpsi_{\rm R}^{\rm N} - \tpsi_{\rm R}^{\rm W} - \tpsi_{\rm R}^{\rm E}
+ \tpsi_{\rm R}^{\rm S} - {e Q h \over 2 r} (2 \tpsi_{\rm I}^{\rm N} -
\tpsi_{\rm I}^{\rm W} - \tpsi_{\rm I}^{\rm E}) +
{\tV h^2 \over 4} \tpsi_{\rm R}^{\rm N} = 0, \label{diffa} \\
\tpsi_{\rm I}^{\rm N} - \tpsi_{\rm I}^{\rm W} - \tpsi_{\rm I}^{\rm E}
+ \tpsi_{\rm I}^{\rm S} + {e Q h \over 2 r} (2 \tpsi_{\rm R}^{\rm N} -
\tpsi_{\rm R}^{\rm W} - \tpsi_{\rm R}^{\rm E}) +
{\tV h^2 \over 4} \tpsi_{\rm I}^{\rm N} = 0, \label{diffb}
\een
where if $\tpsi^{\rm S}$ denotes the value of the 
considered
field at point with
coordinates $(u_0,v_0)$, $\tpsi^{\rm W}$, $\tpsi^{\rm E}$, $\tpsi^{\rm
N}$ are the values of the field at $(u_0+h,v_0)$, $(u_0,v_0+h)$,
$(u_0+h,v_0+h)$, respectively (see Fig.~\ref{fig1}).
Thus given the initial values of the field on the $u=u_0$ axis and
imposing boundary conditions on the $v=v_0$ axis one may integrate
Eqs. (\ref{diffa}) and (\ref{diffb}) row by row. As an initial
condition we choose the Gaussian impulse of the form
\be
\tpsi = (1+i) \exp \left[ - {(v-v^*)^2 \over \sigma} \right]
\ee
In all calculations presented here we use $v^*=100$ and $\sigma=400$. The
size of the grid ranges from $0$ to 
$0$ to $10^4$ or $4 \times 10^4$
in both $u$ and
$v$ and usually $h$ equals at least $0.1$. Because of the scale
invariance of the problem we choose $2 M = 1$. Unless otherwise noted
we use $Q=0.45$ and $e=0.01$.

\begin{figure}[t]
\begin{center}
\leavevmode
\epsfxsize=220pt
\epsfbox{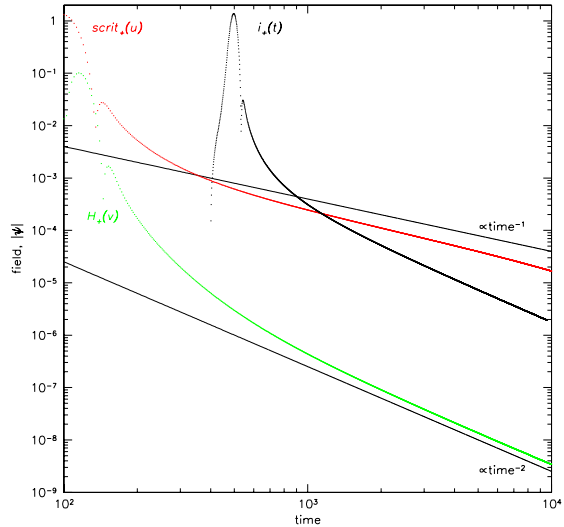}
\end{center}
\caption{Time evolution of the scalar field from the Gaussian initial
data. $i_+$ approximated by $\tpsi(y=400,t)$ represents future
timelike inifinity while $scrit_+$ ($\tpsi(v=4 \times 10^4,u)$) and
$H_+$ ($\tpsi(u=2 \times 10^4,v)$) represent future null infinity and
the future horizon, respectively. Numerically calculated slopes for
these curves are $-2.04$ for $i_+$, $-1.13$ for $scrit_+$, and $-1.99$
for $H_+$. Thin, solid lines have slopes equal to power-law exponents
expected from theoretical predictions.}
\label{fig2}
\end{figure}

\begin{figure}[t]
\begin{center}
\leavevmode
\epsfxsize=220pt
\epsfbox{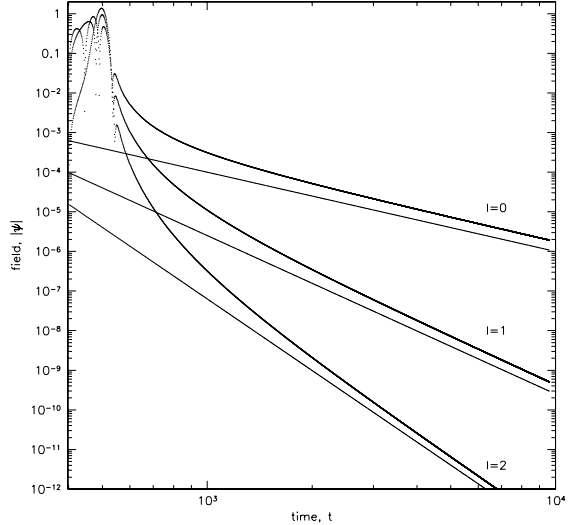}
\end{center}
\caption{Field evolution on future timelike infinity, $i_+$, for different
values of the parameter $l=0,1,2$. Curves have slopes of $-2.04$,
$-4.07$, and $-6.06$, respectively. Thin solid lines indicate power-
law exponents predicted from analytical investigations. }
\label{fig3}
\end{figure}

\begin{figure}[t]
\begin{center}
\leavevmode
\epsfxsize=220pt
\epsfbox{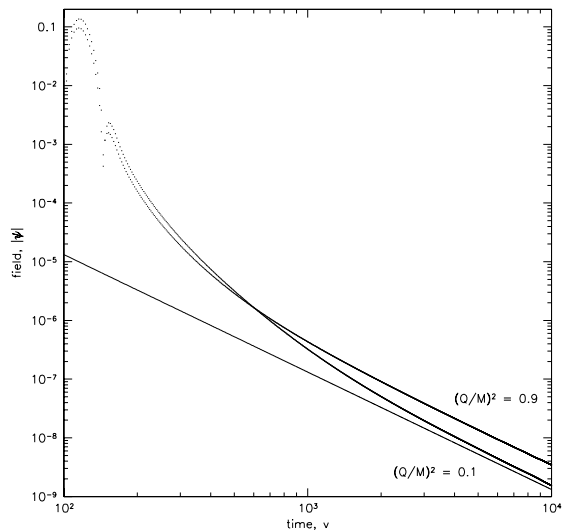}
\end{center}
\caption{Field evolution on future timelike infinity for different
values of $(Q/M)^2$. Power-law exponents of the late-time tails are 
$- 2.07$ for $(Q/M)^2 = 0.1$ and $- 1.98$ for $(Q/M)^2 = 0.9$.}
\label{fig4}
\end{figure}

In Fig.~\ref{fig2} we present time evolution of the field on three
different hypersurfaces: future timelike infinity $i_+$, future null
infinity $scri_+$, and the future horizon $H_+$. In the numerical
experiments these lines are approximated by the values of the field
$\tpsi(y=400,t)$, $\tpsi(v=4000,u)$, and 
$\tpsi(u=2 \times 10^4,v)$
respectively. Calculations are done for $l=0$. After a short period of
a quasinormal ringing the dominance of power-law tails becomes
prominent. The solid lines indicate power law exponents $-1$ and $-2$
expected from the theoretical predictions of Sec.~\ref{sec1}. As can be
seen the agreement between these predictions and the experiment is
excellent. 
The highest discrepancy for $scrit_+$ results from the fact
that theoretical assumption $v \gg u$ is not well maintained at the
end of the line.
In Fig.~\ref{fig3} we examine the dependence of power-law exponents
on the $l$-number. 
We studied  the behaviour of the
field on future timelike infinity
$i_+$ for $l=0,1,2$. As in the previous example the thin solid lines
indicate the theoretically predicted values of power-law exponents from
Eq.~(\ref{ip}). Once again the late time the power-law tails of the field
converge to those theoretically predicted.
In order
to check the late time behaviour of the field depends on the charge $Q$
we performed the calculation for two values of $(Q/M)^2$ 
ratios $0.1$ and $0.9$. 
For the Gaussian initial data with $l = 0$.
The
results are presented in Fig.~\ref{fig4}. 
Once again the theoretical predictions
are in agreement with the numerical simulation.
Fig.~\ref{fig4} envisages the evidence for the
existence of power-law tails on the outer horizon of dilaton black hole
for generic perturbations being of a great importance for the {\it mass
inflation} scenario. 
For the first time the existence of power-law tails on the outer
horizon of RN black hole was revealed in \cite{gun} studing a scalar field
perturbations and also confirmed for the case of a massless charged
scalar fields on RN background \cite{pir1,pir3}.

\section{Conclusions}
In our work we studied the asymptotic late-time behaviour of a massless 
charged scalar field in the background of dilaton black hole
being the spherical solution of the so-called low-energy string theory.
We confined the existence of oscillatory inverse power-law tails in a
collapsing spacetime along asymptotic regions of future timelike
infinity, future null infinity and along the outer horizon of the
considered black hole.
\par
On a null grid $(u, v)$
we integrated numerically the linearized charged scalar field Eqs.
and confirmed the existence of the power-law
tails. One also treated the dependence on the $l$-number,
they converged to the results theoretically predicted. The numerical
results confirmed our analytical conjecture that a charged hair decayed
slower than a neutral one. The same statement was also revealed in RN
case \cite{pir1,pir3}.
The oscillatory inverse power-law tails along the outer horizon
of dilaton black hole suggests the occurence of {\it mass inflation}
along the Cauchy horizon of a dynamically formed dilaton black hole.
It will be interesting to study the fully nonlinear gravitational collapse 
using a different scheme based on
double null coordinates \cite{dn} which allows one to begin with
regular initial conditions and continue the evolution all the way inside
black hole. We hope to return to this problem elsewhere.
Another problem for investigations is the problem of late-time 
behaviour of self-interacting scalar fields in the spacetime of 
dilaton black hole and
the higher order string corrections modifying the initial action.
\cite{pr}.

\vspace{3cm}
\noindent
{\bf Acknowledgements:}\\
RM was partially supported by the Long Term
Astrophysics grant NASA-NAG-6337 and NSF grant No. AST-9529175.
MR was supported in part by KBN grant 2 P03B 093 18.


\end{document}